# Superradiance in an external classical nonresonant field under the Stark interaction with a vacuum field

A.M. Basharov
National Research Center "Kurchatov Institute,"
Moscow, 123182 Russia
Moscow Institute of Physics and Technology (Technical University),
Dolgoprudnyi, Moscow oblast, 141701 Russia
e-mail: basharov@gmail.com

A.I. Trubilko
St. Petersburg State University of State Fire Service of Emercom of Russia,
St. Petersburg, 196105 Russia
trubilko.andrey@gmail.com

Generalized with respect to the Stark interaction of atoms with a vacuum field of zero photon density, Dicke's model is used to describe the Raman superradiance of a localized ensemble of identical atoms in a coherent non-resonant light wave. It is shown that at a certain critical number of atoms in the ensemble the stabilization effect of the excited state of the ensemble relative to collective atomic decay is possible, with superradiance of the atomic ensemble being suppressed. When the number of atoms is close to the critical value, superradiance exhibits features that are opposite to the effects of the conventional superradiance in the pulse delay time. The identified features depend on the intensity of an external coherent field as well as make it possible to assess whether the Stark interaction of an ensemble with a photon-free vacuum electromagnetic field is significant or not.

## 1. Introduction

In creating quantum memory, there has been considerable interest in optical effects under two-photon and Raman excitation and decay of qubits (two-level particles) [1, 2]. In addition, in case of the multi-qubit ensemble, superradiance is a strong undesirable side effect [3], including Raman superradiance [4,5], which both increases the collective decay rate of multi-qubit systems and significantly reduces the time interval to manipulate quantum memory. Further below we will discuss a general mechanism of such problems, which allows "suppressing" Raman superradiance with an optimal choice of the number of qubits in an array of quantum memory.

Despite its wide use, the notion "superradiance" [6], according to different approaches, means anyway the dynamics of an ensemble of a sufficiently large number of identical particles. Among a variety of models for this situation, we will only review those ones regarding an ensemble of identical quantum particles localized in an area with dimensions much smaller than the characteristic wavelengths. For the first time (along with the notion of superradiance) it appeared in Dicke's work [7]. Although the model of localized quantum particles can be easily implemented for superradiance in the microwave range, the simulation of this situation leads to an additional effort within optical and infrared ranges. Generally, a model is implemented when the particles are localized in a single-mode cavity [8]. Thus, in order to simulate the cases of interest, it is sufficient to use the cavity mode with a fairly wide band, i.e. sufficiently low-Q cavity. In this context, particular mention should be made of the experimental studies performed in the dynamics of collective emission of atomic ensembles of $^{87}$Rb [9] and $^{87}$Sr [10] located into the resonator. These research studies are focused on creating superradiant lasers with ultranarrow linewidths used in optical clocks. An experimental method proposed in [11] is aimed at creating a metastable state of an ensemble of Ba atoms using an intercombination superradiative effect. Acting on an ensemble by Stokes laser pumping, the detection of a superradiance pulse indicates the preparation of the desired state of atoms. To describe the results of the experiment, the authors make use of the well-known two-level model, noting that the calculated delay time of the SR pulse differs from the detected one.

In the transition to the Raman resonance Dicke's model can be generalized as follows. Initially, Dicke's model described a localized ensemble of identical two-level particles interacting with a broadband quantized vacuum field in an electric-dipole manner. In this case, the interaction corresponds to the approximation, which is commonly referred to as the rotating wave approximation [12,13], or the resonance approximation. The transition from the excited atomic level to the ground level occurs with the photon emission (Fig.1a) and is a first-order process in the interaction energy.

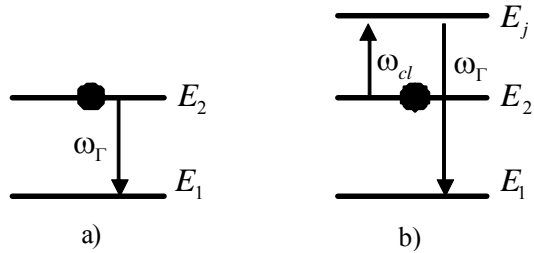

Fig. 1. The scheme of the resonant energy levels of a quantum particle, making a conventional one-quantum transition from the upper level $E_2$ to the lower level $E_1$ with the emission of a photon (a) and a two-quantum transition $E_2 \to E_1$ with the absorption of a photon $\omega_{cl}$ and emission of a photon $\omega_\Gamma$. Nonresonant atomic levels are designated by $E_j$. The transition $E_2 \to E_1$ for the case of a) is optically allowed, whereas in the case of b) it is optically forbidden.

To describe Raman resonance, the notion of a two-level particle and the approximation of a rotating wave is also used, and the transition from the excited level is due to the simultaneous absorption of a Stokes photon from the scattered field and emission of an anti-Stokes photon (Fig.1b). Unlike the initial Dicke's model, there occur second-order processes of the atomic interaction with an electromagnetic field. At the same time, nonresonant atomic levels and their importance are not generally taken into account. In particular, the terms responsible for the shifts of the resonance levels are not usually considered in the Hamiltonian of the system. The Lamb shift is taken into account by renormalizing the resonant transition frequency, and the level shifts corresponding to the high-frequency Stark effect in classical fields are even unaddressed in broadband quantized fields. As a result, theoretical studies of superradiance, including the Raman one, do not take into account either the effect of the Stark interaction of a quantum system with an external broadband quantized field, or the stabilization effect of the excited states of the atomic system [14] relative to its collective radiative decay due to the Stark interaction. This effect of suppressing collective spontaneous emission by the Stark interaction with the vacuum field can be very useful in quantum memory devices.

The work [14] highlights particular importance of the Stark interaction of a quantum particle with a photon-free vacuum (broadband quantized) electromagnetic field in the process of its usual spontaneous decay. In the case of Dicke superradiance, the Stark interaction is always small compared to the resonance interaction energy, but the Stark interaction operator has other algebraic properties compared to a one-quantum resonant transition operator, since, mathematically, the Stark interaction in the Markov approximation is believed to be a quantum counting process. The property of counting becomes apparent in a special competition between two processes in which the excitation of a quantum particle can participate. The particle can pass into the lower energy state upon emitting a real quantum. This process is described by the quantum annihilation process, which determines the quantum Brownian motion as well. However, an excited particle can also participate in processes with virtual photons, which form Stark shifts in energy levels and maintain the particle excitation. As a consequence, weak reemission forming the Stark interaction has proven to "integrate" into a collective radiative decay and suppress superradiance [14] with a certain number of atoms in the ensemble.

In the case of Raman superradiance, the Stark operator of the interaction of atoms with a quantized vacuum electromagnetic field is of the same order as the Raman transition operator [15] and, therefore, it should be taken into account in correct calculations. Meanwhile, the fact has not been referred to in the aforementioned and other works on Raman superradiance. It is usually believed that for the case of a vacuum field with a zero photon density the average value of the Stark interaction operator is zero and it can be omitted in deriving the kinetic equation [16].

In order to highlight the role played by the effect of suppressing collective decay in Raman superradiance in an easier way, this paper considers Raman superradiance of a localized ensemble of identical particles in an external classical coherent electromagnetic field. The classical field acts on atoms along with a quantized broadband vacuum field of zero photon density. The absorption of a photon from the classical field leads to the transition of excited atoms to the ground state with the emission of an anti-Stokes photon into the vacuum field. (see Fig.1b). In this case, both the excited and ground levels are characterized by the same parity, so that the transition between them belongs to optically forbidden transitions. On the one hand, this case is too complicated to analyze as there is a variety of two-quantum relaxation processes of a single atom, some of which are described in [17] without regard to Stark level shifts in a quantized vacuum field, because the classical coherent field was assumed to be sufficiently intense, and the number of atoms is small in [17]. On the other hand, in considering two quantized fields it is also necessary to account for the variety of different processes. Moreover, in the Markov approximation, it is necessary to imply strict limitations to simplify quantum calculations [18].

The present paper has proposed the theory of Raman superradiance of a localized ensemble of identical particles that simultaneously interact with a classical coherent wave and a broadband vacuum electromagnetic field based both on the algebraic perturbation theory and quantum stochastic differential equation (SDE). The algebraic perturbation theory [19–21] is an integral part of the theory of optical interactions with quantum systems; however, it is still not widespread among the experts in nonlinear and quantum optics, as well as quantum SDEs [22–24]. In terms of quantum SDE, evolution of a quantum ensemble and its environment is unitary, and the SDE terms generated by a quantum annihilation process describe a real two-quantum Raman transition from the excited level to the ground one with quantum emission and photon absorption from a coherent wave. A SDE term generated by the counting process describes virtual emission processes and quantum absorption from a quantized electromagnetic field, which are not accompanied by any transitions between quantum atomic levels. The nontrivial Hudson-Parthasarathy algebra [25] for the Ito differentials of the main quantum processes determines the final kinetic equation for the atomic density matrix (master equation), making calculations easier. The paper has presented the derivation of kinetic equations for the density matrix of collectively decaying identical quantum particles under conditions of Raman resonance with the emission of anti-Stokes photons (Fig. 1b) and with regard to the Stark interaction with the quantized broadband electromagnetic field. As in [17], the field is believed to be classical if its photons are absorbed during a two-quantum transition with the emission of an anti-Stokes photon. Discovered in [14], the stabilization effect of excitation in a sufficiently dense ensemble of quantum particles can also be shown to occur in two-quantum Raman superradiance. Application of quantum SDE methods resulted in a kinetic equation for a two-quantum spontaneously radiating atomic ensemble, which turned out to be different from the well-known works on Raman superradiance [1-5] and others by the so-called non-Wiener factor [26,27 ], which manifested some oscillating dependence on the number of particles in an atomic ensemble. This causes the stabilization effect of excitation in a fairly dense ensemble of quantum particles. The stabilization effects of excitation and suppression of superradiance are demonstrated by the simplest analytical solutions of the obtained kinetic equation. The critical values of the number of collectively decaying atoms were obtained, in which the spontaneous emission of an ensemble of excited atoms was completely suppressed. It is shown that under conditions of incomplete suppression of the collective decay of an atomic ensemble, the delay time of a Raman superradiance pulse substantially depends on the difference in the values of Stark constant operating levels. Under certain conditions, a fully excited atomic ensemble may not have a delay, and a semi-excited ensemble of atoms acquires it. Therefore, the time delay of a superradiance pulse can serve as a hallmark of Stark interaction importance in Raman superradiance. Consequently, in comparison with superradiance theories at one-photon resonance [14,26-28], which takes into account the suppression effect of collective decay, our research results have shown that the stabilization effect of excited states found in [14] by the Stark interaction is quite common in a localized ensemble of identical atoms and can be used in quantum memory devices.

## 2. Formulation of the problem

Let an immobile atom, due to the nature of the excitation, can populate only such levels as ground $E_1$ and excited ones $E_2$, which have no other energy levels between themselves. Let these levels be characterized by the same parity, so that the transition $E_2 \to E_1$ is optically forbidden. Then, an excited atom, interacting with the surrounding vacuum electromagnetic field, can pass into the ground state only with the help of various two-quantum processes, for example, by emitting two photons with $\omega_1$ and $\omega_2$ frequencies, such that $\omega_1 + \omega_2 \approx (E_2 - E_1)/\hbar$. An atom can also absorb a photon with $\omega_1$ frequency and emit a photon of a higher frequency $\omega_2 \approx \omega_1 + (E_2 - E_1)/\hbar$. There can also exist other scenarios for the transition into the ground state. Only some particular conditions, such as the ratio of parameters of two-quantum processes, can affect the choice of a possible way of implementation. For simplicity, let us study only one typical scheme of a two-quantum transition, namely, the one shown in Fig.1b. For an atom to be able to absorb a photon $\omega_1$ from the vacuum field, the photon density of the vacuum field should be different from zero at a frequency $\omega_1$. This field will be treated as a classical field with the central carrier frequency $\omega_{cl}$. For the sake of simplicity and certainty, we will consider the following model, whose solution would provide a more definitive assessment for the spontaneous emission of atoms during two-quantum transitions.

Described above, the ensemble of identical immobile atoms interacts only with a quantized electromagnetic field with a central frequency $\omega_\Gamma$ and a classical electromagnetic field with a carrier frequency $\omega_{cl}$ in the electric dipole approximation, with $\omega_\Gamma - \omega_{cl} \approx (E_2 - E_1)/\hbar$. All atoms in the number of $N_a$ are localized in a volume whose dimensions are much smaller than the characteristic wavelengths of electromagnetic fields. The initial Hamiltonian of the system

$$H^{Ini} = H^A + H^F + H^{Int} \tag{1}$$

consists of Hamiltonians' sum of isolated atoms $H^A$, the Hamiltonian of a quantized electromagnetic field $H^F$, and the operator of atomic interaction with electromagnetic fields $H^{Int} = H^{Int-q} + H^{Int-cl}$, the interaction with a quantized field being described by $H^{Int-q}$ and the interaction with a classical field - by $H^{Int-cl}$. These operators have a usual form

$$H^A = \sum_{i,j} E_j |E_j>^{(i)} <E_j|^{(i)},\ H^F = \sum_\omega \hbar\omega b_\omega^+ b_\omega,\ H^{Int-q} = \int d\omega \Gamma_\omega (b_\omega^+ + b_\omega) \sum_{i,kj} d_{kj} |E_k>^{(i)} <E_j|^{(i)},$$

$$\tag{2}$$

$$H^{Int-cl} = -(\mathscr{E} e^{-i\omega_{cl} t} + c.c.) \sum_{i,kj} d_{kj} |E_k>^{(i)} <E_j|^{(i)},\ \sum_j |E_j>^{(i)} <E_j|^{(i)} = 1^{(i)},\ <E_j|^{(i)} E_k>^{(i)} = \delta_{jk},$$

where quantum nondegenerate states $|E_j>$ with energy $E_j$ define the atom, $d_{kj} = <E_k|d|E_j>$ denote matrix elements of the atom dipole moment operator $d = \sum_{kj} d_{kj} |E_k><E_j|$. We consider that all atomic levels are characterized by some certain parity, so that $<E_k|d|E_k>=0$. The superscript of the state vectors marks the state space of the $i$-th atom, and $i$-summation is performed over all atoms of the ensemble involved. The photons' annihilation and creation operators of ω frequency are expressed by the values $b_\omega$ and $b_\omega^+$: $[b_\omega, b_{\omega'}^+] = \delta_{\omega\omega'}$. $\mathscr{E}$ denotes the classical field amplitude, with letters *c.c.* expressing a term complex conjugate of the previous one. Recoil effects and polarization features are neglected. The dipole – dipole interaction of identical atoms is neglected, as in the case of obtaining the main results of the conventional theory of superradiance [6, 7].

In the interaction representation, the wave vector $|\Psi>$ of the system "atom + quantized electromagnetic field" satisfies the Schrödinger equation

$$i\hbar \frac{d}{dt}|\Psi(t)> = H^{Int}(t)|\Psi(t)>, \qquad (3)$$

$$|\Psi(t)> = \exp(i(H^A + H^F)t/\hbar)|\Psi>,$$

$$H^{Int}(t) = \exp(i(H^A + H^F)t/\hbar) H^{Int} \exp(-i(H^A + H^F)t/\hbar) = H^{Int-q}(t) + H^{Int-cl}(t),$$

$$H^{Int-q}(t) = \int d\omega \Gamma_\omega (b_\omega^+ e^{i\omega t} + b_\omega e^{-i\omega t}) \sum_{i,kj} d_{kj} |E_k>^{(i)} <E_j|^{(i)} e^{i\omega_{kj} t},$$

$$H^{Int-cl}(t) = -(\mathscr{E}e^{-i\omega_{cl}t} + c.c.) \sum_{i,kj} d_{kj} |E_k>^{(i)} <E_j|^{(i)} e^{i\omega_{kj}t}, \quad \omega_{kj} = (E_k - E_j)/\hbar,$$

$|\Psi>$ is the wave function of the entire system in the Schrödinger representation.

Furthermore, in order to correctly formulate a quantum SDE, we introduce the effective Hamiltonian representation [20] in making a unitary transformation

$$|\widetilde{\Psi}(t)> = U(t)|\Psi(t)>,$$

$$i\hbar \frac{d}{dt}|\widetilde{\Psi}(t)> = \widetilde{H}^{Int}(t)|\widetilde{\Psi}(t)>, \qquad (4)$$

$$\widetilde{H}^{Int}(t) = U(t)H^{Int}(t)U(t)^+ - i\hbar U(t)\tfrac{d}{dt}U(t)^+. \qquad (5)$$

We express the unitary operator $U(t)$ by the Hermitian operator

$$U(t) = e^{-iS(t)}, \quad S(t)^+ = S(t), \qquad (6)$$

so as to use the Baker-Hausdorf formula for the arbitrary operator $O$

$$e^{-iS} O e^{iS} = O + \frac{(-i)}{1!}[S,O] + \frac{(-i)^2}{2!}[S,[S,O]] + \frac{(-i)^3}{3!}[S,[S,[S,O]]] + \dots$$

Transformed operators (5) and $S(t)$ are expanded in a series of interaction constants

$$S(t) = S^{(10)}(t) + S^{(01)}(t) + S^{(11)}(t) + \dots, \qquad (7)$$

$$\widetilde{H}^{Int}(t) = \widetilde{H}^{Int(10)}(t) + \widetilde{H}^{Int(01)}(t) + \widetilde{H}^{Int(11)}(t) + \widetilde{H}^{Int(20)}(t) + \widetilde{H}^{Int(02)}(t) + \dots,$$

where the left index of each pair indicates the order of expansion in a coupling constant with the classical field, and the right index with the quantized field.

We plug (7) in (5) with account of (6), the Baker-Hausdorff formula and the expressions for $H^{Int-q}(t)$ and $H^{Int-cl}(t)$. Equating expressions of the same order of smallness and leaving in $\widetilde{H}^{Int}(t)$ only those terms that do not have factors varying rapidly in time (refer to [14,18,20] for more detail), we obtain

$$\widetilde{H}^{Int(10)}(t) = \widetilde{H}^{Int(01)}(t) = 0,$$

$$\widetilde{H}^{Int(11)}(t) = \int d\omega \Gamma_\omega b_\omega \mathscr{E}^* e^{-i(\omega - \omega_{cl} - \omega_{21})t} \tfrac{1}{2}\{\Pi_{21}(\omega) + \Pi_{21}(-\omega_{cl})\} R_+ + H.c.,$$

$$\widetilde{H}^{Int(20)}(t) = H^{Stark-cl}(t), \quad \widetilde{H}^{Int(02)}(t) = H^{Stark-q}(t) + H^{Lamb},$$

$$H^{Stark-q}(t) = \int d\omega \Gamma_\omega \int d\omega' \Gamma_{\omega'} b_\omega^+ b_{\omega'} e^{-i(\omega'-\omega)t} \{\Pi_+(\omega,\omega')\frac{N_a}{2} + \Pi_-(\omega,\omega')R_3\},$$

$$H^{Lamb} = \int d\omega \sum_{i,kj} \Gamma_\omega^2 \frac{|d_{kj}|^2}{\hbar(\omega_{kj}-\omega)} |E_k>^{(i)}<E_k|^{(i)}, \quad H^{Stark-cl}(t) = |\mathscr{E}|^2 \sum_{i,k} \Pi_k(\omega_{cl}) |E_k>^{(i)}<E_k|^{(i)},$$

$$\Pi_\pm(\omega,\omega') = \tfrac{1}{2}\{\Pi_2(\omega) + \Pi_2(\omega') \pm (\Pi_1(\omega) + \Pi_1(\omega'))\}.$$

Here the standard parameters of the theory of optical resonance [9]

$$\Pi_{nm}(\omega) = \sum_j \frac{d_{nj} d_{jm}}{\hbar}\left(\frac{1}{\omega_{jn}+\omega} + \frac{1}{\omega_{jm}-\omega}\right) = \Pi_{mn}^*(-\omega), \quad \Pi_k(\omega) = \sum_{jk} \frac{|d_{kj}|^2}{\hbar}\left(\frac{1}{\omega_{kj}+\omega} + \frac{1}{\omega_{kj}-\omega}\right)$$

and collective atomic operators are introduced

$$R_3 = \tfrac{1}{2}\sum_i (|E_2>^{(i)}<E_2|^{(i)} - |E_1>^{(i)}<E_1|^{(i)}), \quad R_- = \sum_i |E_1>^{(i)}<E_2|^{(i)}, \quad R_+ = \sum_i |E_2>^{(i)}<E_1|^{(i)},$$

with the latter obeying the commutation relations of *su*(2) algebra

$$[R_3, R_\pm] = \pm R_\pm, \quad [R_+, R_-] = 2R_3.$$

When deriving $\widetilde{H}^{Int}(t)$, the operator of the excitations exchange between identical particles was ignored. Depending on the properties of a classical field, the operator of Stark interaction $H^{Stark-cl}(t)$ with a classical field can be represented in different forms. In this paper, we regard the classical field as being coherent with the constant amplitude $\mathscr{E}$, therefore, it is convenient to exclude the operators of Stark interaction with the classical field $H^{Stark-cl}(t)$ and the Lamb shift operator $H^{Lamb}$ from the Schrödinger equations (5) by a corresponding unitary transformation, including, in fact, the Stark level shift in the classical field and the Lamb shift in determining the two-quantum transition frequency $\omega_{21}$.

## 3. Basic equations in the Markov approximation

We write the main equation (4) with the effective Hamiltonian $\widetilde{H}^{Int}(t) = \widetilde{H}^{Int(11)}(t) + H^{Stark-q}(t)$ in a dimensionless form and the Markov approximation. We take a quantity $\tau = (\omega'_{21}+\omega_{cl})t$ as a dimensionless time, $\nu = \omega/(\omega'_{21}+\omega_{cl})$ as a dimensionless frequency. Here $\omega'_{21}$ is the frequency of a two-quantum transition $E_2 \to E_1$ with account of the Stark effect in the classical field and the Lamb shift of levels. The value $\omega'_{21}+\omega_{cl}$ is the central frequency $\omega_\Gamma$ of the broadband quantized electromagnetic field, we assume that $\omega_\Gamma = \omega'_{21}+\omega_{cl}$. We express the Schrödinger equation as

$$i\frac{d}{d\tau}|\Psi(\tau)> = (H^{Tr}(\tau) + H^{St}(\tau))|\Psi(\tau)>, \quad (8)$$

$$H^{Tr}(\tau) = \frac{1}{\sqrt{2\pi}}\int_0^\infty d\nu\, b_\nu^+ e^{i(\nu-1)\tau}\chi^R(\nu)R_- + \frac{1}{\sqrt{2\pi}}\int_0^\infty d\nu\, b_\nu e^{-i(\nu-1)\tau}\chi^R(\nu)R_+, \quad (9)$$

$$H^{St}(\tau) = \frac{1}{2\pi}\int_0^\infty d\nu\, b_\nu^+ e^{i(\nu-1)\tau}\int_0^\infty d\nu'\, b_{\nu'} e^{-i(\nu'-1)\tau}\{\eta_+(\nu,\nu')\frac{N_a}{2} + \eta_-(\nu,\nu')R_3\}, \quad (10)$$

$$|\Psi(\tau)> = \exp\{i(H^{Lamb} + H^{Stark-cl}(\tau/\omega_\Gamma))\tau/(\hbar\omega_\Gamma)\}|\widetilde{\Psi}(\tau/\omega_\Gamma)>, \quad [b_\nu, b_{\nu'}^+] = \delta(\nu-\nu').$$

The introduced dimensionless parameters $\chi^R(\nu)$ and $\eta_\pm(\nu,\nu')$ are related to the dimensional values by the ratios

$$\chi^R(\nu) = \frac{\sqrt{2}\omega_\Gamma D_{21}}{\mu c^{3/2}\sqrt{\hbar}}\nu, \quad \eta_\pm(\nu,\nu') = \chi^R(\nu)\chi^R(\nu')\frac{\Pi_\pm(\omega_\Gamma\nu, \omega_\Gamma\nu')}{D_{21}^2/(\hbar\omega_\Gamma)}, \quad D_{21} = \tfrac{1}{2}\{\Pi_{21}(\omega) + \Pi_{21}(-\omega_{cl})\}\mathscr{E}^*.$$

Here $\mu = \sqrt{3}$ is the correction factor [14], which allows taking into account various field geometries; the value $D_{21}$ is a matrix element of the effective operator of the dipole moment of the optically forbidden transition $E_2 \to E_1$ [20]. For the sake of simplicity, we regard $D_{21}$ and $\mathcal{E}$ as real values.

It should be noted that operators $H^{Tr}(\tau)$ and $H^{St}(\tau)$ are dimensionless forms of operators $\tilde{H}^{Int(11)}(t)$ and $H^{Stark-q}(t)$. We have intentionally written the main equations (8) - (10) in the form similar to that one of the equations in [14]. In contrast to describing various physical problems, the main differences are as follows. In [14], the values $H^{Tr}(\tau)$ and $H^{St}(\tau)$ have different orders over the coupling constants with the electromagnetic field. For our case of the Raman resonance these values are of the same order. The parameter of the operator $H^{Tr}(\tau)$ depends on the intensity of the external coherent field, that allows controlling processes of collective interaction of the atomic ensemble. As the intensity of the external field increases, the influence of the difference in the parameters of the Stark interaction $\eta_{\pm}(\nu,\nu')$ on the process of collective decay of the atomic ensemble grows. This difference is discussed in a more detail in the following section.

The initial condition is the factorized wave function $|\Psi_0\rangle = |\Psi_0^A\rangle \otimes |\Psi_0^F\rangle$, where $|\Psi_0^A\rangle$ is the wave function of an atomic subsystem, and $|\Psi_0^F\rangle$ is the wave function of the electromagnetic field's state, with

$$\langle\Psi_0^F|b_\nu^+ b_{\nu'}|\Psi_0^F\rangle = 0, \ \langle\Psi_0^F|b_\nu b_{\nu'}^+|\Psi_0^F\rangle = \delta(\nu-\nu'), \ \langle\Psi_0^F|b_\nu b_{\nu'}|\Psi_0^F\rangle = \langle\Psi_0^F|b_\nu^+ b_{\nu'}^+|\Psi_0^F\rangle = 0,$$
(11)

$$\langle\Psi_0^F|b_\nu|\Psi_0^F\rangle = \langle\Psi_0^F|b_\nu^+|\Psi_0^F\rangle = 0.$$

Let us assume the solution of equation (8) using the evolution operator $U(\tau)$:

$$|\Psi(\tau)\rangle = U(\tau)|\Psi_0\rangle, \ \frac{d}{d\tau}U(\tau) = -i(H^{Tr}(\tau)+H^{St}(\tau))U(\tau), \ U(0)=I,$$
(12)

which is expressed as the $\tilde{T}$ exponent ($I$ is the unit operator)

$$U(t) = I + (-i)\int_0^\tau (H^{Tr}(\tau')+H^{St}(\tau'))d\tau' + (-i)^2 \int_0^\tau \int_0^{\tau'} (H^{Tr}(\tau')+H^{St}(\tau'))(H^{Tr}(\tau'')+H^{St}(\tau''))d\tau' d\tau'' + \ldots =$$

$$= \tilde{T}\exp\left(-i\int_0^\tau (H^{Tr}(\tau')+H^{St}(\tau'))d\tau'\right).$$

(13)

The following assumptions introduce the Markov approximation [23]:

$$b(\tau) = \frac{1}{\sqrt{2\pi}}\int_{-\infty}^\infty d\nu e^{-i(\nu-1)\tau}b_\nu, \ b^+(\tau) = \frac{1}{\sqrt{2\pi}}\int_{-\infty}^\infty d\nu e^{i(\nu-1)\tau}b_\nu^+$$

$$\chi^R(\nu) = const = \chi^R(1) \equiv \chi^R, \ \eta_{\pm}(\nu,\nu') = const = \eta_{\pm}(1,1) \equiv \eta_{\pm},$$
(14)

$$H^{Tr}(\tau)d\tau = \chi^R R_+ dB(\tau) + \chi^R R_- dB^+(\tau), \ H^{St}(\tau)d\tau = (\eta_+ \frac{N_a}{2} + \eta_- R_3)d\Lambda(\tau),$$
(15)

where

$$dB(\tau) = B(\tau+d\tau) - B(\tau), \ dB^+(\tau) = B^+(\tau+d\tau) - B^+(\tau), \ d\Lambda(\tau) = \Lambda(\tau+d\tau) - \Lambda(\tau),$$
(16)

$$B(\tau) = \int_0^\tau d\tau' b(\tau'), \ B^+(\tau) = \int_0^\tau d\tau' b^+(\tau'), \ \Lambda(\tau) = \int_0^\tau d\tau' b^+(\tau')b(\tau'),$$
(17)

with

$$[b(\tau),b^+(\tau')] = \delta(\tau-\tau'), \ [B(\tau),B^+(\tau)] = \tau, \ [B(\tau_1),B^+(\tau_2)] = \int_0^{\tau_1} d\tau' \int_0^{\tau_2} d\tau'' \delta(\tau'-\tau'') = \min(\tau_1,\tau_2).$$

The integrals in (13) are to be interpreted in the Ito sense [22,23]. Roughly speaking, the differentials (16) are the Ito increments of the quantum Wiener $B(\tau)$ and Poisson $\Lambda(\tau)$ processes (more precisely, they determine quantum Wiener and Poisson processes using the formulas [29]). In neglecting the Stark interaction $\eta_\pm = 0$, the outlined approach is in line with the well-known descriptions of the processes of spontaneous emission by methods for quantum SDEs [23].

The efficiency and ease of the quantum SDE method is determined by the fact that the increments (16) satisfy the Hudson-Parthasarathy algebra [25]:

$$d\Lambda(\tau)d\Lambda(\tau) = d\Lambda(\tau), \quad d\Lambda(\tau)dB^+(\tau) = dB^+(\tau), \quad dB(\tau)d\Lambda(\tau) = dB(\tau), \quad dB(\tau)dB^+(\tau) = d\tau. \quad (18)$$

$$d\Lambda(\tau)dB(\tau) = d\Lambda(\tau)d\tau = dB^+(\tau)d\Lambda(\tau) = dB^+(\tau)d\tau = dB(\tau)d\tau = 0.$$

In order to obtain a quantum SDE (in the Ito sense) for the evolution operator $U(\tau)$ (instead of equation (12), which is uncertain under the Markov approximation), the increment $dU(\tau)$ defined as $dU(\tau) = U(\tau + d\tau) - U(\tau)$ should be considered. Taking into account the representation (6) similar to [14], we have

$$dU(\tau) = \{\exp(-i(\chi^R R_+ dB(\tau) + \chi^R R_- dB^+(\tau) + (\eta_+ \frac{N_a}{2} + \eta_- R_3)d\Lambda(\tau))) - 1\}U(\tau).$$

Expanding the exponent in a series and using the Hudson-Parthasarathy algebra, we obtain a quantum SDE for the evolution operator in the form

$$dU(\tau) = A_0 d\tau U(\tau) + A_+ dB(\tau)U(\tau) + A_- dB^+(\tau)U(\tau) + A_\Lambda d\Lambda(\tau)U(\tau), \quad (19)$$

where the operator's functions

$$A_0 = (\chi^R)^2 R_+ \frac{e^{-i(\eta_+ \frac{N_a}{2} + \eta_- R_3)} - 1 + i(\eta_+ \frac{N_a}{2} + \eta_- R_3)}{(\eta_+ \frac{N_a}{2} + \eta_- R_3)^2} R_-, \quad A_- = \frac{e^{-i(\eta_+ \frac{N_a}{2} + \eta_- R_3)} - 1}{\eta_+ \frac{N_a}{2} + \eta_- R_3} \chi^R R_-,$$

$$A_+ = \chi^R R_+ \frac{e^{-i(\eta_+ \frac{N_a}{2} + \eta_- R_3)} - 1}{\eta_+ \frac{N_a}{2} + \eta_- R_3}, \quad A_\Lambda = e^{-i(\eta_+ \frac{N_a}{2} + \eta_- R_3)} - 1$$

are to be interpreted as Taylor's series of the corresponding functions with the argument $x$, resulting from the replacement $(\eta_+ \frac{N_a}{2} + \eta_- R_3) \to x$ with the subsequent reverse replacement $x \to (\eta_+ \frac{N_a}{2} + \eta_- R_3)$.

The equation for the density matrix $\rho(\tau) = U(\tau)|\Psi_0\rangle\langle\Psi_0|U^+(\tau)$ of the given system, which consists of identical atoms and a broadband electromagnetic field, is derived by calculating the increment $d\rho(\tau) = \rho(\tau + d\tau) - \rho(\tau)$. Using (19) and the Hudson-Parthasarathy algebra (18) leads to

$$d\rho(\tau) = A_0 d\tau \rho(\tau) + A_+ dB(\tau)\rho(\tau) + A_- dB^+(\tau)\rho(\tau) + A_\Lambda d\Lambda(\tau)\rho(\tau) +$$
$$+ \rho(\tau)A_0^+ d\tau + \rho(\tau)dB^+(\tau)A_+^+ + \rho(\tau)dB(\tau)A_-^+ + \rho(\tau)d\Lambda(\tau)A_\Lambda^+ +$$
$$+ A_+ dB(\tau)\rho(\tau)dB^+(\tau)A_+^+ + A_+ dB(\tau)\rho(\tau)dB(\tau)A_-^+ + A_+ dB(\tau)\rho(\tau)d\Lambda(\tau)A_\Lambda^+ +$$
$$+ A_- dB^+(\tau)\rho(\tau)dB^+(\tau)A_+^+ + A_- dB^+(\tau)\rho(t)dB(\tau)A_-^+ + A_- dB^+(\tau)\rho(\tau)d\Lambda(\tau)A_\Lambda^+ +$$
$$+ A_\Lambda d\Lambda(\tau)\rho(\tau)dB^+(\tau)A_+^+ + A_\Lambda d\Lambda(\tau)\rho(\tau)dB(\tau)A_-^+ + A_\Lambda d\Lambda(\tau)\rho(\tau)d\Lambda(\tau)A_\Lambda^+.$$

Hence the atomic density matrix $\rho^A(\tau)$ is obtained by tracing over the field variables $\rho^A(\tau) = Tr_F \rho(\tau)$:

$$\frac{d\rho^A}{d\tau} = (\chi^R)^2 a_-^{NL}(\eta_+,\eta_+,R_3) R_- \rho^A R_+ a_+^{NL}(\eta_+,\eta_+,R_3) -$$
$$-\frac{(\chi^R)^2}{2}\{R_+\{a_0^{NL}(\eta_+,\eta_+,R_3) - ia_s^{NL}(\eta_+,\eta_+,R_3)\}R_-\rho^A + \rho^A R_+\{a_0^{NL}(\eta_+,\eta_+,R_3) + ia_s^{NL}(\eta_+,\eta_+,R_3)\}R_-\}.$$
(20)

Non-Wiener operators are introduced here

$$a_0^{NL}(\eta_+,\eta_+,R_3) = 2\frac{1-\cos(\eta_+\frac{N_a}{2}+\eta_- R_3)}{(\eta_+\frac{N_a}{2}+\eta_- R_3)^2}, \quad a_s^{NL}(\eta_+,\eta_+,R_3) = 2\frac{\eta_+\frac{N_a}{2}+\eta_- R_3 - \sin(\eta_+\frac{N_a}{2}+\eta_- R_3)}{(\eta_+\frac{N_a}{2}+\eta_- R_3)^2},$$

$$a_\pm^{NL}(\eta_+,\eta_+,R_3) = \frac{\cos(\eta_+\frac{N_a}{2}+\eta_- R_3)-1}{\eta_+\frac{N_a}{2}+\eta_- R_3} \pm i\frac{\sin(\eta_+\frac{N_a}{2}+\eta_- R_3)}{\eta_+\frac{N}{2}+\eta_- R_3},$$

and $a_+^{NL}(\eta_+,\eta_+,R_3)a_-^{NL}(\eta_+,\eta_+,R_3) = a_0^{NL}(\eta_+,\eta_+,R_3)$. In the process of derivation, the relation $Tr_F(\rho(\tau)dB(\tau)) = Tr_F(\rho(\tau)dB^+(\tau)) = Tr_F(\rho(\tau)d\Lambda(\tau)) = 0$ was used.

Equation (20) describes the dynamics of an ensemble of identical atoms that collectively interact with a common (vacuum) broadband electromagnetic field with a zero photon density and are concentrated in a small volume with dimensions much smaller than the wavelength of the emitted photons. Following [14,18,28], equations similar to (19) and (20) determined by the presence/allowance for a counting (Poisson) process increment $d\Lambda(\tau)$, as well as the dynamics of quantum systems described by them, are called non-Wiener equations and non-Wiener (generalized Langevin) dynamics (in contrast to the Langevin dynamics determined by the presence/allowance for increments of Wiener processes $dB(\tau)$ and $dB^+(\tau)$ only).

According to the outlined approach, a counting process increment $d\Lambda(\tau)$ describes the Stark interaction of an ensemble of identical atoms with a vacuum electromagnetic field. It was first taken into account in describing atomic dynamics in [30], whose results follow from equation (13) for a single atom $N_a = 1$ and a simplified model of the Stark interaction $\eta_+ = 0$. Due to the relation $d\Lambda(\tau)d\Lambda(\tau) = d\Lambda(\tau)$ in a quantum SDE method, it is possible to automatically sum up the entire series of perturbation theory that occurs in alternative approaches if one makes an attempt to take into account the Stark interaction. Consequently, even a small Stark interaction of a single atom can prove to significantly affect the collective spontaneous emission of an ensemble of a sufficient number of identical atoms. Further below it is illustrated by the collective decay of a singly excited atomic ensemble.

**4. Collective emission of a singly excited atomic ensemble**

Let a singularly excited ensemble symmetrized by permutations of $N_a$ atoms be prepared at some initial moment of time, which can be described by the wave function

$$|\Psi_0\rangle = \frac{1}{\sqrt{N}}(|E_2\rangle^{(1)}|E_1\rangle^{(2)}...|E_1\rangle^{(N)} + |E_1\rangle^{(1)}|E_2\rangle^{(2)}...|E_1\rangle^{(N)} + ...|E_1\rangle^{(1)}|E_1\rangle^{(2)}...|E_2\rangle^{(N)}). \quad (21)$$

The presented state is related to the one of W class. Its collective two-quantum decay from an excited level under the action of a classical field will be described by the kinetic equation (20), whose solution can be easily obtained in the Dicke function basis $|r,m\rangle$. These are eigenfunctions of the Casimir operator $R^2 = \frac{1}{2}(R^+R^- + R^-R^+) + \frac{1}{4}R_3^2$ and inversion $R_3$:

$$R^2|r,m\rangle = r(r+1)|r,m\rangle, \quad R_3|r,m\rangle = m|r,m\rangle, \quad (22)$$

which form a (2r + 1) - dimensional representation of the su(2) algebra, where generators are operators $R_3$ and $R^\pm$: $R^\pm |r,m\rangle = \sqrt{(r \mp m)(r \pm m + 1)}|r, m \pm 1\rangle$. The reference state (21) of an atomic system in this basis is described by the vector $|r, -r+1\rangle$, where $r = \frac{N_a}{2}$, whose collective two-quantum decay with a photon absorption is determined by a kinetic equation for matrix element relaxation $\langle r, -r+1|\rho^A|r, -r+1\rangle = \rho^A_{-r+1,-r+1}$, which follows from (20) and has a simple form

$$\frac{d\rho^A_{-r+1,-r+1}}{d\tau} = -4(\chi^R)^2 r \frac{1-\cos(r\eta_+ - r\eta_-)}{(r\eta_+ - r\eta_-)^2} \rho^A_{-r+1,-r+1}. \qquad (23)$$

Consequently, for non-Wiener dynamics, the collective Raman decay of the given state is of an exponential nature

$$\rho^A_{-r+1,-r+1}(\tau) = \exp(-4(\chi^R)^2 r \frac{1-\cos(r\eta_+ - r\eta_-)}{(r\eta_+ - r\eta_-)^2} \tau).$$

Compared with the case of the Wiener dynamics of the atomic system, the decay rate appears to be modulated by a periodic nonlinear function

$$f = 8 \frac{(1-\cos(\frac{N_a}{2}(\eta_+ - \eta_-)))}{(N_a(\eta_+ - \eta_-))^2},$$

generated by an account of the Stark interaction. The interaction constant $\chi^R$ is proportional to the amplitude of an external classical field involved in the two-quantum transition of an atom, which is considered to be a real value. By varying it, one can exert influence on the rate of the two-quantum state decay by changing its magnitude. In contrast to Wiener dynamics, where $\eta_\pm = 0$, and the Raman transition rate is proportional to both the amplitude of the classical field involved in the transition, and the number of atoms $N_a$, an account of the Stark interaction gives rise to a factor as a function $f$ in the equation (23). It is this very function that determines the critical value of the number of atoms $N_*$ in the ensemble for any nonzero difference value $(\eta_+ - \eta_-)$, at which the two-quantum superradiance spontaneous decay of the $W$ state is completely suppressed. This critical number of atoms is determined from the following condition

$$N_* = 4\pi \frac{k}{(\eta_+ - \eta_-)}, \quad k = 1,2,3... \qquad (24)$$

It is to be noted that for the model involved the critical number of atoms, at whose value the collective two-quantum decay of a single-excited ensemble symmetrized in atoms' permutation is completely suppressed, is determined by the Stark interaction parameter of a lower level $\Pi_1$ and not affected by the classical field's amplitude

$$N_* = 2\pi k \frac{3c^3}{\omega_\Gamma^3 v^2 \Pi_1(\omega_\Gamma v)}.$$

Thus, we have showed that in a two-quantum spontaneous decay with an anti-Stokes photon emission symmetrized in permutations of a single-excited atomic ensemble, an account of the Stark interaction of even a single atom under the conditions of its environment with a sufficient number of atoms in the ensemble significantly affects the collective relaxation in the system. For some critical values, cooperative relaxation can be completely suppressed, leading to stabilization of the system's state. At the same time, the intensity of an external classical field involved in the process determines the magnitude of its collective relaxation rate by means of controlling the latter one.

**5. Collective decay of an excited atomic ensemble**

Let us now consider a collective two-quantum decay of a fully excited symmetric state of the atomic ensemble and its semi-excited state, where one half of the atoms is in the excited state, and the other half is in the ground one. Note that in the Dicke basis a fully excited state of the system is described by the function $\left|\frac{N_a}{2},\frac{N_a}{2}\right\rangle$, a semi-excited ensemble by $\left|\frac{N_a}{2},0\right\rangle$, and the ground state is represented by the vector $\left|\frac{N_a}{2},-\frac{N_a}{2}\right\rangle$.

The master equation (20) for the atomic density matrix elements combines its diagonal elements only $\langle r,m|\rho^A|r,m\rangle = \rho^A_{m,m}$, which can be written in the following form

$$\frac{d\rho^A_{m,m}}{d\tau} = -2(\chi^R)^2 g_{m,m-1} C_{m-1} \rho^A_{m,m} + 2(\chi^R)^2 g_{m+1,m} C_m \rho^A_{m+1,m+1}. \quad (25)$$

Here the rate coefficient $(\chi^R)^2$ depends on the amplitude of the classical field that initiates the Raman cooperative emission in the system, along with coefficients $g_{m,m-1} = \langle r,m|R^+|r,m-1\rangle\langle r,m-1|R^-|r,m\rangle = (r+m)(r-m+1)$, also determine collective relaxation within the Wiener dynamics of the system description, whereas the latter takes into account only resonant two-quantum processes of interaction between the atomic system and the vacuum broadband environment. Additional factors, such as

$$C_m = \frac{1-\cos(\eta_+ \frac{N_a}{2} - \eta_- m)}{(\eta_+ \frac{N_a}{2} - \eta_- m)^2}$$

determine a two-photon collective relaxation rate in the non-Wiener dynamics of the atomic system; their origin is related to an account of both resonant and non-resonant interactions in the system.

It is evident that the excited collective state of the system passes through a whole set $|r,m\rangle$ of states of the "Dicke ladder" in the course of its evolution. For those of them which satisfy the condition

$$\eta_+ \frac{N_*}{2} + \eta_- m = 2\pi k, \quad k = 1,2,3,4.... \quad (26)$$

the collective two-quantum relaxation rate appears to be fully suppressed, its value is equal to zero, and the states themselves are stabilized. In this case their stabilization conditions depend not only on the number of atoms in the system, and also on the position of the state vector on the Dicke ladder, given the difference in the Stark interaction constant of the working levels $\eta_- \neq 0$.

Under the Wiener dynamics of a fully excited symmetrized atomic ensemble of a large number of atoms $N_a \gg 1$, the superradiance pulse intensity is determined by the well-known relation

$$I(\tau) = q\frac{N_a^2}{4}\gamma_0 \text{sech}^2(\gamma_0 \frac{N_a}{2}(\tau-\tau_0)),$$

where $q$ is the geometric factor, $\gamma_0 = 2(\chi^R)^2$ is the ensemble's relaxation rate $N_a$, $\tau_0 = (\gamma_0 N_a)^{-1}\ln(N_a)$ is its time delay. The latter is calculated by determining the average emission time $\langle\tau\rangle$ by the $n$ anti-Stokes photon emission during two-quantum transition in the field of a classical wave from the excited state. For the general case, this quantity is determined by the sum

$$\langle\tau\rangle = \sum_{\frac{N}{2}-n-\frac{1}{2}}^{\frac{N}{2}}(\gamma_0 g_{m,m-1})^{-1}.$$

The Raman pulse of superradiance of the semi-excited state of the atomic system occurs without time delay, since the upper limit of the reduced amount in this case is zero.

In the general case, the intensity of cooperative anti-Stokes emission in a two-quantum transition in the field of a classical wave can be defined as the decrease in the energy of the atomic ensemble $I(\tau) = -q \frac{d}{d\tau} Sp(\hbar \omega_\Gamma R_3 \rho^A)$ and calculated as the following sum

$$I(\tau) = \sum_{m=-\frac{N_a}{2}}^{m=\frac{N_a}{2}} q \gamma_0 g_{m,m-1} C_{m-1} \rho_{m,m} = q_0 \sum_{-\frac{N_a}{2}}^{\frac{N_a}{2}} G_{m,m-1} \rho_{m,m}.$$

For the non-Wiener dynamics, the function $G_{m,m-1}$ and the coefficients $g_{m,m-1}$ are determined by the function $C_{m-1}$, which is defined by both the number of $N_a$ atoms in the system and the eigenvalue $m$ of the inversion operator for different values of Stark interaction constants of the working levels. This fact does not allow for an analytical calculation, so we will present the results of numerical calculations.

We will be treating the pulse of collective two-quantum emission of a fully excited symmetric atomic ensemble $\left|\frac{N_a}{2}, \frac{N_a}{2}\right\rangle$, as well as the pulse from a semi-excited atomic ensemble, whose state $\left|\frac{N_a}{2}, 0\right\rangle$, depending on the intensity of a classical field leading to the Raman transition in the system with an anti-Stokes photon emission. Since the nature of emission essentially depends on the critical value $N_*$ of the number of atoms in the system, which, as shown in [14], should be no less than a hundred, we will discuss pulses of this emission far from the critical value of the number of atoms, at the values of which the collective two-quantum emission is fully suppressed. This critical value is determined in terms of the relation (26), which is the argument of the function $C_{m-1}$, where the product of the Stark interaction parameter of the working levels and the number of atoms are important. It is to be noted that when the basic kinetic equation (20) is derived, the following relation between the interaction constants $\eta_\pm \ll \chi^R \ll 1$ should be satisfied, which leads to small values of the Stark interaction. Therefore, all the curves are presented taking into account the renormalization, where $N_a = 8$, and the parameter values $\eta_\pm$ are chosen as fractions of $\pi$ and are indicated in the figures. Moreover, in these figures the intensity magnitude as the square of the real amplitude of the classical pump field is represented by a dimensionless parameter corresponding to its ratio to the intensity of the classical field taken per unit, with the latter determining the time scale, being a factor in the interaction constant $(\chi_0^R)^2$ and dimensionless time $\tau$. We omitted the superscript $R$ of the interaction constant to emphasize the universality of the results presented in the figures. These time dependences are inherent in any model of a localized ensemble of two-level particles that takes into account the Stark interaction with a vacuum electromagnetic field with a zero photon density.

Let us discuss the intensity of Raman superradiance pulse from a fully excited atomic ensemble. Figure 2 provides graphs of a collective two-quantum emission pulse during its Wiener dynamics. An increase in the intensity of the classical coherent field performing a resonant two-quantum transition with the emission of an anti-Stokes photon leads to an increase in the intensity value at the maximum and a decrease in the pulse delay time.

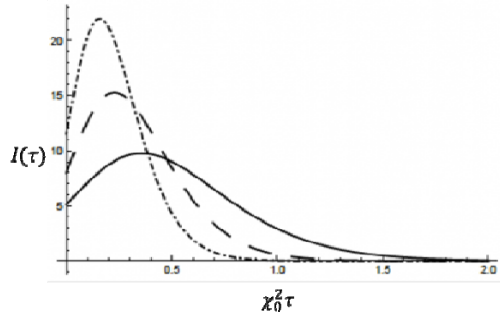

Fig.2. The intensity of a two-quantum superradiance pulse from time for the Wiener dynamics of the atomic ensemble. The dimensionless intensity of the classic field: solid curve – 0.64, dotted curve –1.0, dash-dotted curve –1.44.

The dependence applies to a superradiance pulse with the non-Winner dynamics far from the critical value of the number of atoms in the ensemble, which corresponds to the full suppression of the system's cooperative relaxation. Thus, in Fig.3a,b, curves of superradiant pulses emitted by a fully excited atomic system are presented both for the number of atoms smaller (a) and larger (b) than the first critical value, when the Stark interaction constants of the working levels are identical $\eta_- = 0$.

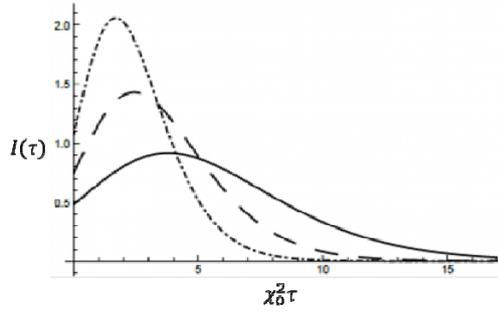

Fig.3,a. The intensity of the two-quantum superradiance pulse from time for the non-Winner dynamics of a fully excited atomic ensemble for the number of atoms in the ensemble smaller than the first critical value $\eta_+ = \pi/2 - 0.4$. The parameters for the Stark interaction of the working levels are identical $\eta_- = 0$. The dimensionless intensity of the classic field: solid curve – 0.64, dotted curve –1.0, dash-dotted curve –1.44.

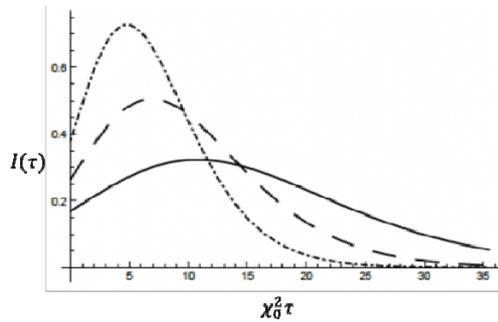

Fig.3,b. The intensity of the two-quantum superradiance pulse from time for the non-Winner dynamics of a fully excited atomic ensemble for the number of atoms in the ensemble larger than the first critical value $\eta_+ = \pi/2 + 0.4$. The parameters for the Stark interaction of the working levels are identical $\eta_- = 0$. The dimensionless intensity of the classic field: solid curve – 0.64, dotted curve –1.0, dash-dotted curve –1.44.

When the values of the Stark parameters are different $\eta_- \neq 0$, the dependence on the amplitude of the classical coherent wave in two-quantum radiation remains unchanged. This is given in Figure 4,a for the number of atoms in the ensemble of smaller than the critical ones, and in Figure 4,b for the values of the number of atoms in the ensemble of larger than the first critical one. In the latter case, there seems to be no sign of time delay for superradiance pulses, which is a distinctive feature of the non-Wiener dynamics (under the conditions of a difference in the Stark interaction constants of working levels).

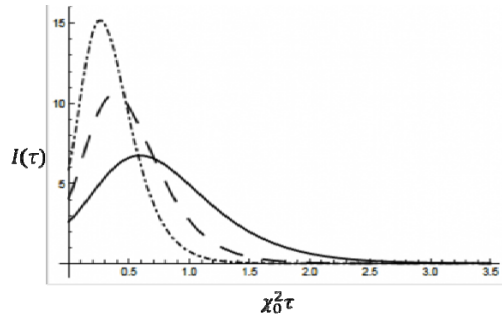

Fig.4,a. The intensity of the two-quantum superradiance pulse from time for the non-Winner dynamics of a fully excited atomic ensemble for the number of atoms in the ensemble smaller than the first critical value $\eta_+ = \pi/8$. The parameters for the Stark interaction of the working levels are different $\eta_- = \pi/8$. The dimensionless intensity of the classic field: solid curve – 0.64, dotted curve –1.0, dash-dotted curve –1.44.

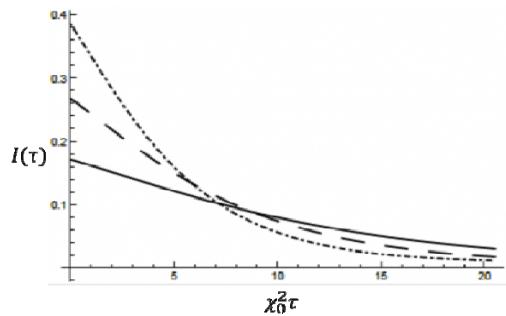

Fig.4,b. The intensity of the two-quantum superradiance pulse from time for the non-Winner dynamics of a fully excited atomic ensemble for the number of atoms in the ensemble larger than the first critical value $\eta_+ = \pi/4 + 0.6$. The parameters for the Stark interaction of the working levels are different $\eta_- = \pi/4$. The dimensionless intensity of the classic field: solid curve – 0.64, dotted curve –1.0, dash-dotted curve –1.44.

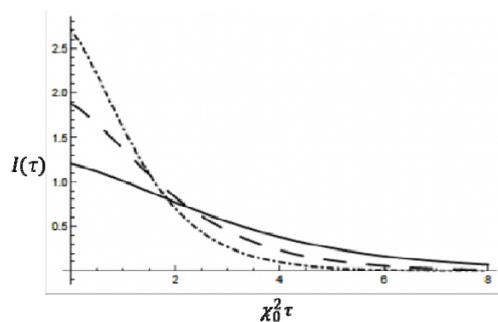

Fig.5,a. The intensity of the two-quantum superradiance pulse from time for the non-Winner dynamics of a semi-excited atomic ensemble for the number of atoms in the ensemble smaller

than the first critical value $\eta_+ = \pi/2 - 0.4$. The parameters for the Stark interaction of the working levels are identical $\eta_- = 0$. The dimensionless intensity of the classic field: solid curve– 0.64, dotted curve –1.0, dash-dotted curve –1.44.

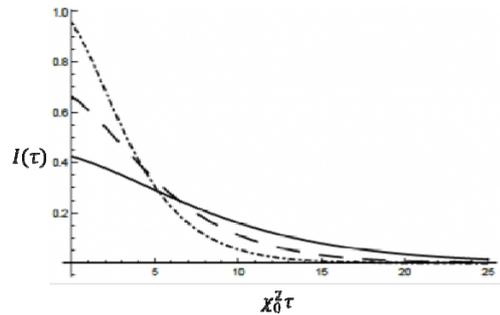

Fig.5,b. The intensity of the two-quantum superradiance pulse from time for the non-Winner dynamics of a semi-excited atomic ensemble for the number of atoms in the ensemble larger than the first critical value $\eta_+ = \pi/2 + 0.4$. The parameters for the Stark interaction of the working levels are identical $\eta_- = 0$. The dimensionless intensity of the classic field: solid curve – 0.64, dotted curve –1.0, dash-dotted curve –1.44.

The cooperative emission by a semi-excited atomic ensemble in a two-quantum Raman process is presented in Fig.5 and Fig.6. With a change in the intensity of the classical field, the superradiance pulse dependence remains unchanged - an increase in its value leads to an increase in the highest pulse intensity far from the critical value of the number of atoms in the ensemble. This is given in Fig. 5,a for the number of atoms in the ensemble of smaller than the critical one, and in Figure 5,b for the values of the number of atoms in the ensemble larger than the first critical one (under the conditions of an equality in the Stark interaction constants of working levels). It is to be noted that there seems to be no sign of any pulse delay as is the case with the non-Wiener dynamics. In the case of different values of Stark interaction constants of the working levels of the atomic system, for the number of ensemble emitters smaller than the critical value, along with the given dependence on the classical field amplitude, there appears a delay in the superradiance pulse, which decreases in magnitude with increasing the external field energy, as shown in Fig. 6, a. Fig. 6, b shows that the pulse delay is not observed when the number of atoms is greater than the critical one.

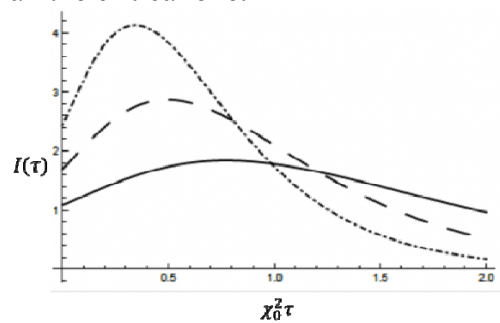

Fig.6,a. The intensity of the two-quantum superradiance pulse from time for the non-Winner dynamics of a semi-excited atomic ensemble for the number of atoms in the ensemble smaller than the first critical value $\eta_+ = \pi/4 + 0.6$. The parameters for the Stark interaction of the working levels are different $\eta_- = \pi/4$. The dimensionless intensity of the classic field: solid curve – 0.64, dotted curve –1.0, dash-dotted curve –1.44.

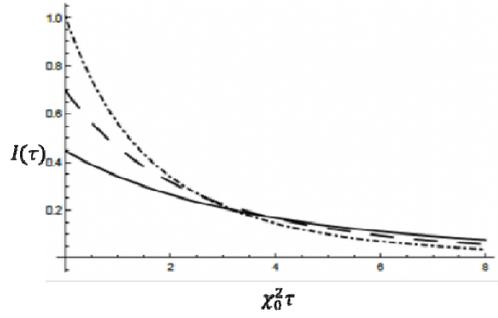

Fig.6,b. The intensity of the two-quantum superradiance pulse from time for the non-Winner dynamics of a semi-excited atomic ensemble for the number of atoms in the ensemble larger than the first critical value $\eta_+ = \pi/4 + 1.4$. The parameters for the Stark interaction of the working levels are different $\eta_+ = \pi/4$. The dimensionless intensity of the classic field: solid curve – 0.64, dotted curve –1.0, dash-dotted curve –1.44.

## 6. Conclusion

The paper describes the model of Raman superradiance of a localized ensemble in an external coherent nonresonant field with an account of the Stark interaction of atoms with a vacuum field of zero photon density. The model was built on the basis of the algebraic perturbation theory up to the second order inclusively and a quantum SDE. Despite fact that the model clearly deals only with two working levels under Raman resonance conditions, the model parameters take into account all atomic energy levels. The quantum SDE of the model differs from those of the commonly used ones in quantum optics [23] by a counting quantum random process. In addition to our works [14, 18, 26–28, 30] in quantum optics, the counting process was accounted only in [31–35] under other resonance conditions and for a single atom. Meanwhile, calculations with an account of the counting process have been made for a long time [24, 25]. Nevertheless, the role of the counting process has been particularly clear and has increased only in an ensemble of a sufficient number of atoms [14]. As the present paper shows this has resulted in the suppression of Raman superradiance by Stark interaction with a certain number of atoms of the ensemble. At the same time, by varying the intensity of the coherent field for the case when the Stark parameters of working levels differ from each other, it becomes possible to experimentally answer the question whether the Stark interaction with the vacuum field is significant or not. The answer to the question lies in the delay times of the Raman superradiance of a fully excited atomic ensemble and in its semi-excited state. In contrast to the conventional superradiance, the Raman superradiance delay under the conditions of the stabilization effect of the excited state of an atomic ensemble (with respect to the collective atomic decay) takes place at a semi-excited state and is not encountered in the case of a fully excited atomic ensemble. As a result, these conditions could be much in demand in quantum memory devices when the Stark interaction is essential and there occurs the stabilization effect of the excited states of the atomic ensemble with respect to the collective atomic decay.

References

1. Z. Zhang, C.H. Lee, R. Kumar, K.J. Arnold, S.J. Masson, A.L. Grimsmo, A.S. Parkins and M.D. Barrett. Phys.Rev. A **97**, 043858 (2018).
2. D.-S. Ding, et al. Nature Photon. **9**, 332–338 (2015).
3. M.S. Mendes et al. New J. Phys. **15** 075030 (2013)
4. T. Wang, and S.F. Yelin, Phys.Rev. A **72**, 043804 (2005).

5. Y. Yoshikawa, Y. Torii and T. Kuga. Phys.Rev.Lett. **94**, 083602 (2005)
6. R.H. Dicke, Phys. Rev. **93**, 99 (1954).
7. VI.V. Kocharovsky, V.V. Zheleznyakov, E.R. Kocharovskaya and V.V. Kocharovsky, Physics–Uspekhi, **60**, 345 (2017).
8. L. Ortiz-Gutierrez et al. Phys.Rev.Lett. **120**, 083603 (2018).
9. J.G. Bohnet et al, Phys.Rev.Lett. **109**, 253602 (2012).
10. M.A. Norcia et al, Sci. Adv. **2**, e1601231 (2016).
11. C. Ohae et al, J. Phys. Soc.Japan **83**, 044301 (2014).
12. W.H. Louisell, *Quantum Statistical Properties of Radiation*, Wiley, New York, 1973.
13. L. Mandel and E. Wolf, *Optical Coherence and Quantum Optics*, Cambridge University Press, Cambridge, 1995.
14. A.M. Basharov, Phys.Rev. A **84**, 013801 (2011).
15. A.I. Maimistov and A.M. Basharov, *Nonlinear optical waves*, Dordrecht: Kluwer Academic, 1999.
16. K. Blum. *Density Matrix Theory and Applications*. Springer, 1996.
17. A.M. Basharov, Sov. Phys. JETP **75**, 611 (1992).
18. A.I. Trubilko and A.M. Basharov. JETP **126**, 604 (2018).
19. A.M. Basharov, A.I. Maimistov and E.A. Manykin, Sov.Phys.JETP **57**, 282 (1983).
20. A.I. Maimistov and A.M. Basharov, *Nonlinear optical waves*, Kluwer Academic, Dordrecht, 1999.
21. V.N. Bogaevski and A. Povzner. *Algebraic Methods in Nonlinear Perturbation Theory*, Springer, Berlin, 1991.
22. C.W. Gardiner and M.J. Collet, Phys.Rev.A **31**, 3761 (1985).
23. C.W. Gardiner and P. Zoller, *Quantum noise* (Springer-Verlag, Berlin, 2004).
24. A. Holevo. Statistical structure of quantum theory. Springer, Berlin, 2001.
25. R.L. Hudson and K.R. Parthasarathy, Comm.Math.Phys. **93**, 301 (1984).
26. A.M.Basharov, Phys.Lett. A **375**, 2249 (2011)
27. A.M.Basharov, Phys.Let. A, **376**, 1881 (2012).
28. A.I.Trubilko and A.M.Basharov. JETP Lett. **109**, (2019).
29. V.P. Belavkin, Theor.Math.Phys. **110**, 46 (1997).
30. A.M.Basharov, Phys.Lett. A **375**, 2249 (2011).
31. A. Barchielli, Phys.Rev.A **34**, 1642 (1986).
32. A. Barchielli and N. Pero. J. Opt. B: Quantum Semiclass. Opt. **4**, 272 (2002).
33. B.Q. Baragiola, R.L. Cook, A.M. Branczyk and J. Combes, Phys.Rev. A **86**, 013811 (2012).
34. A. Dabrowska, G. Sarbicki and D. Chruscinski, Phys. Rev. A **96**, 053819 (2017).
35. B.Q. Baragiola and J. Combes, Phys. Rev. A **96**, 023819 (2017).